\documentclass[12pt,a4paper]{article}

\usepackage{amsmath,amssymb,amsthm,geometry}
\usepackage{mathrsfs}
\usepackage{hyperref}
\usepackage[table]{xcolor}   
\usepackage{graphicx}   
\usepackage{rotating}
\usepackage{amsmath}
\usepackage{cite}

\usepackage{authblk}

\newcommand{\bb}[1]{\mathbb{#1}}

\newcommand{\Sthree}{\texorpdfstring{$S_3$}{S3}}

\title{Electroweak Structure and Three Fermion Generations in Clifford Algebra with \Sthree\ Family Symmetry}

\author{Niels Gresnigt\\
\small{Department of Physics, Xi’an Jiaotong-Liverpool University, 111 Ren’ai Rd., Suzhou 215123, P.R. China}\\
\small{\texttt{niels.gresnigt@xjtlu.edu.cn}}}
\date{} 

\begin{document}
\maketitle

\begin{abstract}
We construct an explicit algebraic realisation of three fermion generations within a single Clifford algebra, transforming under the full Standard Model $SU(3)_C\times SU(2)_L\times U(1)_Y$ gauge group, in which an intrinsic $S_3$ family symmetry permutes three algebraically distinguished but gauge-equivalent fermion sectors without replicating the gauge bosons. Fermionic states are represented by minimal left ideals of the complex Clifford algebra $\bb{C}\ell(10)$, while the three-generation structure arises from an embedded discrete $S_3$ symmetry acting on the space of algebraic spinors. The Standard Model gauge generators are identified as elements commuting with this $S_3$ action and act on physical states via the adjoint (commutator) action. The resulting spectrum reproduces the correct Standard Model quantum numbers for three linearly independent generations of fermions.
\end{abstract}

\section{Introduction}

The Standard Model (SM) of particle physics represents a triumph of mathematical modeling, offering a highly accurate description of elementary matter. Despite its impressive agreement with experimental data, it remains limited in its ability to fully address several fundamental questions. These include among others:
\begin{enumerate}
\item  What foundational principle, if any, gives rise to the gauge group of the SM?
\vspace{-10pt}
\item Why are specific irreducible representations of Lie groups associated with particle multiplets, while others do not exhibit this correspondence?
\vspace{-10pt}
\item What is the underlying origin of the three fermion generations observed in the SM?
\end{enumerate}

In this work, building on our earlier algebraic constructions \cite{Gillard2019,Gresnigt2023,gourlay2024algebraic}\footnote{See also \cite{gillard2019c,gresnigt2019sedenions,gresnigt2023toward,gresnigt2023sedenion}.}, we construct an explicit algebraic realisation of three linearly independent fermion generations transforming under the full SM gauge group $SU(3)_C\times SU(2)_L\times U(1)_Y$. The construction is formulated within the complex Clifford algebra $\bb{C}\ell(10)$ and incorporates an embedded discrete $S_3$ symmetry, interpreted as a family symmetry acting on the space of algebraic spinors. Crucially, the SM gauge generators are required to commute with this $S_3$ action, ensuring that the gauge sector remains generation-independent and is not replicated.

At the technical level, fermionic states are realised as elements of selected minimal left ideals of $\bb{C}\ell(10)$, while gauge transformations act via the adjoint (commutator) action $[\,\cdot\,,\cdot\,]$. The colour and electroweak gauge algebras are identified as distinguished subalgebras preserving a suitable Witt basis, and the action of $S_3$ generates three linearly independent but gauge-equivalent copies of the fermionic spectrum. This structure provides a purely algebraic mechanism for three generations that is compatible with the full SM gauge symmetry.

The remainder of this Introduction places the present work in the context of earlier algebraic approaches based on division algebras and Clifford algebras. We then summarise the key algebraic ingredients needed for the construction, before presenting the explicit realisation of the gauge algebra and the three-generation fermionic spectrum.

In the search for a mathematically minimal framework that accounts for the structure of the SM gauge group, its particle content, and the existence of three generations, there has been growing interest in algebraic approaches to high-energy physics in recent years. In particular, the four normed division algebras ($\bb{R},\bb{C},\bb{H},\bb{O}$), together with the Clifford algebras arising from their left multiplication endomorphisms, have repeatedly shown promise in accounting for the structure and particle content of the SM.

The idea of relating the four normed division algebras to the SM has a long and rich history. Shortly after the discovery of quarks, G\"uynadin and G\"ursey \cite{Gunaydin1973} constructed a model of quark colour symmetry based on the algebra of the split octonions\footnote{It is important to note that the construction is possible only in the split octonions, and not in the octonions themselves.}. Subsequently, a series of papers by Barducci et al, and Casalbuoni and Gatto \cite{Barducci1977,Casalbuoni1979,Casalbuoni1980} explored a unified description of leptons and quarks with internal degrees of freedom in terms of fermionic oscillators (corresponding to a Witt basis of a Clifford algebra). Three fermionic oscillators\footnote{The abstract algebra of three fermionic oscillators has two different realisations, one in terms of split octonions \cite{Gunaydin1973}, the other in terms of the (associative) complex Clifford algebra $\bb{C}\ell(6)$.} are associated with the colour degrees of freedom. The inclusion of electric charge requires a fourth fermionic oscillator (which in \cite{Barducci1977,Casalbuoni1979,Casalbuoni1980} is unrelated to division algebras). Related connections between the octonions and the SM, including their relation to $SU(3)$, have been investigated in \cite{lasenby2024some,marrani2025physics,krasnov2025octonions}.

The early association of the split octonions with quarks in \cite{Gunaydin1973} was expanded upon by Dixon \cite{Dixon1990,Dixon2004,Dixon1994,Dixon2010} who revealed that the mathematical characteristics of a single fermion generation are inherent in $\bb{T}^2$, where $\bb{T}=\bb{R}\otimes\bb{C}\otimes\bb{H}\otimes\bb{O}$, commonly referred to as the Dixon algebra. Dixon treats $\mathbb{T}$ as a spinor, acted upon by the algebra itself, with each possible action corresponding to a linear map. These linear maps generate the so called associative multiplication algebra, in this case the complex Clifford algebra $\bb{C}\ell(10)$, which acts on the spinor space. 

Furey and other have encompassed both bosons and fermions within $\bb{C}\ell(10)$, representing spinors through the minimal left ideals of the algebra \cite{furey2016standard,todorov2021superselection,Gresnigt2020,furey231,furey232,Furey2023Three}. In such approaches, both the gauge groups and fermions are contained within a Clifford algebra, generated from a composition of the four division algebras acting on itself. Although a composition of division algebras need not be associative (nor alternative), the Clifford algebras generated from the algebra acting on itself via its endomorphisms is necessarily associative.

The representation of spinors as minimal left ideals of Clifford algebras dates back to work in the 1930s and 1940s \cite{Juvet1930,Sauter1930,riesz2013clifford}, with their construction via a Witt decomposition reviewed by Ab\l amowicz \cite{Ablamowicz1995}.

Other research has explored models using real Clifford algebras, revealing shared features with division algebra-based models \cite{trayling1999geometric, trayling2001geometric, trayling2004cl,chisholm1996properties,hamilton2023unification}. Group theoretical foundations in particle physics, alongside relevant subgroups of Clifford algebras, are discussed in \cite{wilson2022remarks,wilson2020group,wilson2020subgroups,wilson2021problem}. Division algebraic representations of the exceptional Lie groups $E_6$ and $E_8$ in relation to the SM are explored in \cite{manogue2022octions,dray2024new}. Efforts to integrate gravity into algebraic unification models have also been undertaken by several authors \cite{singh2020trace,singh2021characteristic,perelman2019r,perelman2019c,finster2024causal}.

While these frameworks are mathematically appealing, most successfully describe only a single fermion generation, offering no intrinsic algebraic mechanism for the existence of three. Indeed, despite numerous attempts to account for three generations \cite{manogue1999dimensional,Furey2018a,Furey2014,Dixon1994,Gillard2019,gording2020unified}, division-algebra based constructions to date remain largely limited to a single generation. The same issue plagues the most popular GUTs, such as those based on $SU(5)$, $SO(10)$, and the Pati-Salam model, which are also inherently single-generation models, lacking an explanation for the origin three generations.

Furey endeavours to represent three generations $SU(3)_C\times U(1)_{em}$ gauge symmetry directly from the algebra $\bb{C}\ell(6)$ \cite{Furey2018a,Furey2014}. A similar construction based on $\bb{C}\ell(6)$, which includes an $SU(2)$ gauge symmetry, is given by Gording and Schmidt-May \cite{gording2020unified}. Dixon, on the other hand, characterises three generations using the algebra $\bb{T}^6=\bb{C}\otimes\bb{H}^2\otimes\bb{O}^3$ \cite{Dixon1994}. In the unified theories of \cite{Casalbuoni1980}, an additional $m$ fermionic oscillators are included in order to represent $2^m$ generations (a necessarily even number). However, these additional fermionic oscillators cannot be associated with the division algebras in an obvious way, nor is there any algebraic guidance on what $m$ should be. Other authors have sought to encode three generations within the exceptional Jordan algebra $J_3(\bb{O})$, comprising three-by-three matrices over the octonions $\bb{O}$ \cite{dubois2016exceptional1,dubois2019exceptional2,todorov2018octonions,todorov2018deducing,boyle2020standard2,boyle2020standard}. Each octonion is associated with one generation through the three canonical $J_2(\bb{O})$ subalgebras of $J_3(\bb{O})$.

Several independent approaches to algebraic unification have identified $\bb{C}\ell(8)$ as a natural setting for addressing the existence of three fermion generations. Early work by Silagadze proposed that the triality of
$\mathrm{Spin}(8)$ may underlie the appearance of three generations \cite{silagadze1994so}\footnote{See also \cite{gresnigt2019sedenions,perelman2021jordan,boyle2020standard2,lisi2024c,lisi2007exceptionally} for related discussions on triality in relation to three generations.}.

The author’s own work on sedenions and $\bb{C}\ell(8)$, dates back to 2019 \cite{Gillard2019,gillard2019c}, where the discrete group $S_3$ was identified as a generation symmetry arising from the automorphism structure of the sedenions. Recently, Furey and Hughes considered $\bb{C}\ell(8)$ within a division-algebra and triality-inspired framework \cite{furey2025three}. In a distinct dynamical approach, Quinta derives $\bb{C}\ell(8)$ directly from a Lagrangian formulation \cite{quinta2025spacetime}, with triality again playing a key role and leading to predictive mass relations. Taken together, these developments indicate that $\bb{C}\ell(8)$ arises independently across a range of approaches, reinforcing its relevance for addressing the problem of fermion generations.

The present work builds on a previously proposed algebraic framework accounting for the
existence of three fermion generations
\cite{Gillard2019,Gresnigt2023,gourlay2024algebraic}\footnote{See also
\cite{gillard2019c, gresnigt2019sedenions,gresnigt2023toward,gresnigt2023sedenion}.}.
In this approach, the discrete group $S_3$, arising as a automorphisms of the sedenion
algebra $\bb{S}$ obtained via the Cayley--Dickson construction from the octonions $\bb{O}$,
is identified as a natural candidate for a generation symmetry. Indeed, the automorphism
group of the sedenions is known to be
$\mathrm{Aut}(\bb{S}) \cong G_2 \times S_3$ \cite{brown1967generalized}.

The endomorphisms of the algebra $\bb{C}\otimes\bb{S}$ provide a faithful representation of the complex Clifford algebra $\bb{C}\ell(8)$. In our earlier work, the $S_3$ automorphisms of $\bb{S}$ were embedded into the complex Clifford
algebra $\bb{C}\ell(8)$, yielding an algebraic realisation of three linearly independent
fermion generations transforming under the unbroken gauge group
$SU(3)_C \times U(1)_{em}$. The central idea is to represent physical states as elements of
minimal left ideals of $\bb{C}\ell(8)$, while identifying gauge symmetries as subalgebras
preserving an appropriate Witt basis. Requiring the gauge generators to commute with the
$S_3$ action ensures that the gauge sector is not replicated across generations.

In the present paper, we extend this construction to include the full electroweak sector
$SU(2)_L \times U(1)_Y$, which necessitates enlarging the algebra to $\bb{C}\ell(10)$.
Although the original motivation for the model arose from the sedenion algebra, we work
here directly with $\bb{C}\ell(8)$ and $\bb{C}\ell(10)$, invoking the sedenion origin only
where required.

\section{{$\bb{C}\ell(8)$ basis and its minimal left ideals}}

In previous works \cite{Gresnigt2023,gourlay2024algebraic} we used a $\bb{C}\ell(8)$ basis obtained directly from the hypercomplex algebra of sedenions. Here instead we opt to use a different basis which highlights the underlying geometry and algebraic structure more transparently. In particular we choose a basis in which the primitive idempotents are diagonal. Since $\bb{C}\ell(8)\cong \textrm{Mat}(16,\bb{C})$ is simply, all faithful representations differ by a similarity transformation (inner automorphism), and so all our results are basis independent. 

We define the following $\bb{C}\ell(8)$ basis


\begin{equation}\label{Cl(8)basis}
    \begin{aligned}
        e_1 &= i \sigma_1 \otimes \sigma_1 \otimes \sigma_1 \otimes \sigma_1, \\
        e_2 &= i \sigma_1 \otimes \sigma_1 \otimes \sigma_3 \otimes \bb{I}_2, \\
        e_3 &= -i \sigma_1 \otimes \sigma_1 \otimes \sigma_1 \otimes \sigma_3, \\
        e_4 &= -i \sigma_1 \otimes \sigma_3 \otimes \bb{I}_2 \otimes \bb{I}_2, \\
        e_5 &= -i \sigma_1 \otimes \sigma_1 \otimes \sigma_1 \otimes \sigma_2, \\
        e_6 &= -i \sigma_1 \otimes \sigma_2 \otimes \bb{I}_2 \otimes \bb{I}_2, \\
        e_7 &= i \sigma_1 \otimes \sigma_1 \otimes \sigma_2 \otimes \bb{I}_2, \\
        e_8 &= i\sigma_2 \otimes \bb{I}_2 \otimes \bb{I}_2 \otimes \bb{I}_2,
    \end{aligned}
\end{equation}
which satisfies
\begin{eqnarray}
    e_i^2=-1,\qquad \{e_i,e_j\}=-2\delta_{ij},\qquad i,j=1,...,8.
\end{eqnarray}
The pseudoscalar of the algebra is defined as
\begin{eqnarray}
    \omega_8:=e_1e_2e_3e_4e_5e_6e_7e_8=\sigma_3\otimes \bb{I}_2\otimes \bb{I}_2\otimes \bb{I}_2.
\end{eqnarray}

\subsection{Primitive idempotents in $\bb{C}\ell(8)$}

In order to construct the primitive idempotents and subsequently the minimal left ideals of the algebra, we first define a Witt basis\footnote{See \cite{Ablamowicz1995} for an overview on the construction of minimal left ideals of Clifford algebras.}:
\begin{eqnarray}\label{Cl(8)wittbasis}
        a_j = \frac{1}{2}(-e_j + ie_{j+4}), \qquad a_j^\dagger = \frac{1}{2}(e_j + ie_{j+4}),
\end{eqnarray}
where $j=1,2,3,4$. These satisfy the algebra of of four fermionic oscillators:
\begin{eqnarray}\label{eqn:fermionicalgebra}
\{a_i,a_j\}=0,\qquad \{a_i^\dagger,a_j^\dagger\}=0,\qquad \{a_i,a_j^\dagger\}=\delta_{ij}.
\end{eqnarray}

For each $j=1,2,3,4$ define the simple idempotents
\begin{equation}
\pi_j^{(+)} \;=\; a_j a_j^\dagger \;=\; \tfrac{1}{2}\bigl(1-i\,e_{j}e_{j+4}\bigr),
\qquad
\pi_j^{(-)} \;=\; a_j^\dagger a_j \;=\; \tfrac{1}{2}\bigl(1+i\,e_{j}e_{j+4}\bigr),
\end{equation}
where the image of $1$ in the matrix representation in eqn. (\ref{Cl(8)basis}) is the  $16\times 16$ identity matrix $\bb{I}_{16}$. These form a maximal set of mutually commuting simple idempotents
\begin{eqnarray}
    \pi_j^{(\pm)}\pi_k^{(\pm)}=\pi_k^{(\pm)}\pi_j^{(\pm)}.
\end{eqnarray}
These simple idempotents correspond to rank-8 diagonal matrices\footnote{In fact, given our $\bb{C}\ell(8)$ basis in eqn. (\ref{Cl(8)basis}), the choice of Witt basis in eqn. (\ref{Cl(8)wittbasis}) is the unique choice that gives diagonal simple idempotents.}.

The 16 primitive (rank-1) idempotents are obtained by choosing one simple idempotent from each pair and multiplying:
\begin{equation}
f_{\varepsilon_1\varepsilon_2\varepsilon_3\varepsilon_4}
\;=\;
\pi_1^{(\varepsilon_1)}\pi_2^{(\varepsilon_2)}\pi_3^{(\varepsilon_3)}\pi_4^{(\varepsilon_4)},
\qquad \varepsilon_j\in\{+,-\}.
\end{equation}
These primitive idempotents are mutually orthogonal and complete:
\begin{equation}
f_{\varepsilon} f_{\varepsilon'} = \delta_{\varepsilon_1,\varepsilon'_1}\delta_{\varepsilon_2,\varepsilon'_2}\delta_{\varepsilon_3,\varepsilon'_3}\delta_{\varepsilon_4,\varepsilon'_4} f_{\varepsilon},
\qquad
\sum_{\varepsilon\in\{+,-\}^4} f_{\varepsilon} = \bb{I}_{16},\qquad f_\varepsilon=f_{\varepsilon_1\varepsilon_2\varepsilon_3\varepsilon_4}.
\end{equation}

In the present matrix realisation, these primitive idempotents correspond to matrix units $E_{i,i}$ where $i=1,...,16$. In this paper the focus will be on four $\bb{C}\ell(8)$ primitive idempotents in particular:
\begin{eqnarray}
f_{++++} &=& (a_1 a_1^\dagger)(a_2 a_2^\dagger)(a_3 a_3^\dagger)(a_4 a_4^\dagger) \;=\; E_{1,1}, \\
f_{- - - +} &=& (a_1^\dagger a_1)(a_2^\dagger a_2)(a_3^\dagger a_3)(a_4 a_4^\dagger) \;=\; E_{14,14}, \\
f_{- - - -} &=& (a_1^\dagger a_1)(a_2^\dagger a_2)(a_3^\dagger a_3)(a_4^\dagger a_4) \;=\; E_{6,6}.\\
f_{+++ -} &=& (a_1 a_1^\dagger)(a_2 a_2^\dagger)(a_3 a_3^\dagger)(a_4^\dagger a_4) \;=\; E_{9,9},
\end{eqnarray}

Because,
\begin{eqnarray}
    a_if_{++++}=0,\quad i=1,2,3,4,
\end{eqnarray}
the first minimal left ideal $I_1$ can be written as
\begin{eqnarray}
    I_1=\bb{C}\ell(8)f_{++++}=(r_0 + r_i a_i^\dagger + r_{ij} a_i^\dagger a_j^\dagger + r_{ijk} a_i^\dagger a_j^\dagger a_k^\dagger + r_{1234} a_1^\dagger a_2^\dagger a_3^\dagger a_4^\dagger)f_{++++},
\end{eqnarray}
where $i, j, k \in \{1,2,3,4\}$, $i < j < k$ and $r \in \mathbb{C}$. This minimal left ideal occupies the first column in our representation (\ref{Cl(8)basis}). Each basis element of the minimal ideal is then proportional to some $E_{j,1}$, where $j=1,...,16$. For example, $a_1^{\dagger}a_2^{\dagger}f_{++++}=-E_{4,1}$. One can likewise construct minimal ideals on the primitive idempotents $f_{---+}$, $f_{----}$, and  $f_{+++-}$.

\subsection{Witt-basis preserving symmetries}
Although any transformation of the form
\begin{eqnarray}\label{spinoraction}
  e_i\mapsto e^{i\phi_kg_k}e_ie^{-i\phi_kg_k},\quad \phi_k \in \mathbb{R},\quad g_k \in \mathbb{C}\ell(8), 
\end{eqnarray}
will preserve the anticommutation relations in eqn. (\ref{eqn:fermionicalgebra}), not all such transformations preserve the Witt basis in eqn. (\ref{Cl(8)wittbasis}) or, equivalently, the isotropic subspaces generated by them. In particular, not all elements of $\textrm{Spin}(8)$, whose Lie algebra is generated by the bi-vectors of $\mathbb{C}\ell(8)$, leave these isotropic subspaces invariant. Imposing the additional restrictions that
\begin{eqnarray}
   [g_k, \sum_{i} \kappa_ia_i]=\sum_{j} \mu_ja_j,\quad \textrm{and}\quad [g_k, \sum_{i} \kappa'_ia_i^\dagger]=\sum_{j} \mu'_ja_j^\dagger,
\end{eqnarray}
and that transformations on $a_i^{\dagger}$ ($a_i$) commute with hermitian conjugation $\dagger$;
\begin{eqnarray}
   e^{i\phi_kg_k} a_i^\dagger e^{-i\phi_kg_k}=(e^{-i\phi_kg_k})^\dagger a_i^\dagger (e^{i\phi_kg_k})^\dagger,
\end{eqnarray}
reduces $\textrm{Spin}(8)$ to its maximal subgroup $U(4)=SU(4)\times U(1)$. These unitary symmetries preserve the Witt basis (and hence the minimal left ideals) via the action in eqn. (\ref{spinoraction}). Both the non-abelian $SU(4)$ generators $\Lambda_1-\Lambda_{15}$ and the abelian $U(1)$ generator $N_8$ can be written in terms of the Witt basis, and these are explicitly listed in Appendix \ref{sec:appA}.

Via the commutator (adjoint) action of $SU(4)$, the minimal left ideal $I_1$ transforms as $\Lambda^\bullet \bb{C}^4$, where $\bb{C}^4$ is identified with the four-dimensional complex vector space spanned by the creation operators $a_i^\dagger$ of the Witt basis of $\bb{C}\ell(8)$.
\begin{eqnarray}
I_1 &\cong& \Lambda^0 \mathbb{C}^4 \;\oplus\; \Lambda^1 \mathbb{C}^4 \;\oplus\; \Lambda^2 \mathbb{C}^4 \;\oplus\; \Lambda^3 \mathbb{C}^4 \;\oplus\; \Lambda^4 \mathbb{C}^4 \nonumber\\
&\cong& 1 \;\oplus\; 4 \;\oplus\; 6 \;\oplus\; \overline{4} \;\oplus\; 1.
\end{eqnarray}
In terms of the Witt basis:
\begin{eqnarray}
\Lambda^0 \mathbb{C}^4 &=& \mathrm{span}\{\,f\,\},\\
\Lambda^1 \mathbb{C}^4 &=& \mathrm{span}\{\,a_i^\dagger f\,\}_{i=1}^{4},\\
\Lambda^2 \mathbb{C}^4 &=& \mathrm{span}\{\,a_i^\dagger a_j^\dagger f\,\}_{1\le i<j\le 4},\\
\Lambda^3 \mathbb{C}^4 &=& \mathrm{span}\Big\{\,\tfrac{1}{6}\,\varepsilon_{ijkl}\,a_j^\dagger a_k^\dagger a_l^\dagger f\,\Big\}_{i=1}^{4},\\
\Lambda^4 \mathbb{C}^4 &=& \mathrm{span}\{\,a_1^\dagger a_2^\dagger a_3^\dagger a_4^\dagger f\,\},
\end{eqnarray}
where for brevity we have written $f=f_{++++}$.

Under the  $SU(3)$ subgroup of $SU(4)$, generated by $\Lambda_1,...,\Lambda_8$ the exterior powers restrict as
\begin{eqnarray}
\Lambda^0 \mathbb{C}^4 &\to& 1,\;
\Lambda^1 \mathbb{C}^4 \;\to\; 3\oplus 1,\;
\Lambda^2 \mathbb{C}^4 \;\to\; 3\oplus \overline{3},\;
\Lambda^3 \mathbb{C}^4 \;\to\; \overline{3}\oplus 1,\;
\Lambda^4 \mathbb{C}^4 \;\to\; 1.
\end{eqnarray}
Consequently $I_1$ transforms as
\begin{eqnarray}
I_1\cong 1 \;\oplus\; (3\oplus 1) \;\oplus\; (3\oplus \overline{3}) \;\oplus\; (\overline{3}\oplus 1) \;\oplus\; 1,\qquad (SU(3)).
\end{eqnarray}

 \section{The discrete $S_3$ symmetry in $\bb{C}\ell(8)$}\label{sec:S3}
We will consider the discrete group $S_3$ as a family symmetry in what is to follow. This discrete group corresponds to an automorphism group of the hypercomplex algebra of sedenions, $\bb{S}$, and is related to the triality automorphism of $\textrm{Spin}(8)$ \cite{varadarajan2001spin,brown1967generalized,gourlay2024algebraic}.

$S_3$ is generated by the order three generator $\psi_3$, together with the order two generator $\epsilon$ satisfying $\psi_3^3=\textrm{Id}$, $\epsilon^2=\textrm{Id}$ and $\epsilon\psi_3=\psi_3^2\epsilon$. Here we introduce a realisation of these two generators, and thus $S_3$ within the context of $\bb{C}\ell(8)$. 

To do so, we first introduce the following multivector elements of $\bb{C}\ell(8)$:
\begin{eqnarray}
g_1 &= \frac{1}{2}e_1e_8(B_1-1) ,\\ 
g_{2} &= \frac{1}{2}e_2e_8(B_2-1), \\ 
g_{3} &= \frac{1}{2}e_3e_8(B_3-1), \\ 
g_{4} &=\frac{1}{2}e_4e_8(B_4-1), \\ 
g_{5} &= \frac{1}{2}e_5e_8(B_5-1), \\ 
g_{6} &= \frac{1}{2}e_6e_8(B_6-1), \\ 
g_{7} &= \frac{1}{2}e_7e_8(B_7-1),
\end{eqnarray}
where $B_i$ are the following four-vectors:
\begin{eqnarray}
B_1 &= \; -e_{2345}+e_{2367}+e_{4567},\\
B_2 &= \; e_{1346}+e_{1357}+e_{4567},\\
B_3 &= \; e_{1256}-e_{1247}+e_{4567},\\
B_4 &= \; e_{1256}+e_{1357}+e_{2367},\\
B_5 &= \; -e_{1247}+e_{1346}+e_{2367},\\
B_6 &= \; -e_{1247}+e_{1357}-e_{2345},\\
B_7 &= \; e_{1256}+e_{1346}-e_{2345}.
\end{eqnarray}

Here and throughout, $e_{i_1 i_2 \cdots i_k}$ denotes the Clifford product
$e_{i_1}e_{i_2}\cdots e_{i_k}$ for ordered indices $i_1<i_2<\cdots<i_k$. These multivector elements ultimately arise from the left actions of the sedenion algebra onto itself, with each such action representing a linear transformation. In particular, such left actions by the first eight imaginary sedenion units generate  $\bb{C}\ell(8)$, whereas the left actions of the remaining seven imaginary sedenion units are represented by the specific multivector combinations $g_i$ within $\bb{C}\ell(8)$, as constructed in previous work \cite{Gresnigt2023,gourlay2024algebraic}. Since our focus here is on the Clifford algebraic structure, we take these elements as given and work entirely within $\bb{C}\ell(8)$.


We now define the action of the order-three $S_3$ generator $\psi_3$ on the generating basis of $\bb{C}\ell(8)$ as follows:
\begin{eqnarray}
    \psi_3&:& e_i\mapsto \frac{1}{4}e_i + \frac{\sqrt{3}}{4}g_i - \frac{\sqrt{3}}{4}e_i e_8 - \frac{3}{4}g_i e_8,
\end{eqnarray}
where $i=1,...,7$. Both $e_8$ as well as the identity are fixed by $\psi_3$, consistent with the sedenion automorphisms in \cite{brown1967generalized}. We note that $\psi_3$ does not map one-vectors to one-vectors, and so does not correspond to a $\textrm{Spin}(8)$ transformation. In fact, $\psi_3$ does not preserve the $\mathbb{Z}_2$-grading of $\bb{C}\ell(8)$, and thus mixes even and odd grades. 

Subsequently, we can determine the action of $\psi_3$ on the multivectors $g_i$ using the fact that $\psi_3$ extends to a homomorphism of the algebra
\begin{eqnarray}
    \psi_3&:& g_i\mapsto -\frac{\sqrt{3}}{4}e_i + \frac{1}{4}g_i+ \frac{3}{4}e_i e_8 - \frac{\sqrt{3}}{4}g_i e_8.
\end{eqnarray}

Likewise we define the action of the order-two generator $\epsilon$ on the $\bb{C}\ell(8)$ generating basis:
\begin{eqnarray}
    \epsilon&:& e_i\mapsto -e_i,\quad i=1,...,8.
\end{eqnarray}
This order-two map leaves $g_i$ unchanged
\begin{eqnarray}
    \epsilon&:& g_{i}\mapsto g_{i},\quad i=1,...,7.
\end{eqnarray}

These generators satisfy $\psi_3^3=\textrm{Id}$, $\epsilon^2=\textrm{Id}$ and $\epsilon\psi_3=\psi_3^2\epsilon$, and together they generate $S_3$:
\begin{eqnarray}
\langle \psi_{3},\, \epsilon \mid \psi_{3}^{3} = \mathrm{Id},\ \epsilon^{2} = \mathrm{Id},\ \epsilon\psi_{3} = \psi_{3}^{2}\epsilon \rangle
\;\cong\; S_{3}.
\end{eqnarray}
The order-three generator $\psi_3$ generates the cyclic subgroup $\bb{Z}_3$ of $S_3$.

\subsection{The action of $\epsilon$ and $\psi_3$ on primitive idempotents}
We can now determine the action of the $S_3$ generators on the Witt basis and minimal left ideals of $\bb{C}\ell(8)$. We note that
\begin{eqnarray}
    \epsilon&:& a_i \mapsto -a_i,\quad i=1,2,3,4,
\end{eqnarray}
from which it follows that
\begin{eqnarray}
    \epsilon&:& a_ia_i^{\dagger}\mapsto a_ia_i^{\dagger} \Rightarrow \epsilon: \pi_i^{(\pm)}\mapsto \pi_i^{(\pm)},\quad i=1,2,3,4.
\end{eqnarray}
The primitive idempotents are therefore fixed by $\epsilon$, and each minimal left ideal is closed
under $\epsilon$, which only induces a sign change on the odd-grade terms in the minimal left ideals.

Next, We determine the action of $\psi_3$ on the primitive idempotent $f_{++++}=E_{1,1}$ to be:
\begin{eqnarray}
\psi_{3}(E_{1,1}) &=& 
\tfrac{1}{4} E_{1,1} 
+ \tfrac{3}{4} E_{14,14} 
+ \tfrac{i\sqrt{3}}{4}\,(E_{14,1} - E_{1,14}),\\
\psi_{3}^{2}(E_{1,1}) &=& 
\tfrac{1}{4} E_{1,1} 
+ \tfrac{3}{4} E_{14,14} 
- \tfrac{i\sqrt{3}}{4}\,(E_{14,1} - E_{1,14}),.
\end{eqnarray}
Acting with $\psi_3$ once more returns $E_{1,1}$, as required. It follows immediately that
\begin{eqnarray}
    E_{1,1}+\psi_3(E_{1,1})+\psi_3^2(E_{1,1})=\frac{3}{2}(E_{1,1}+E_{14,14}),
\end{eqnarray}
is $S_3$-invariant.

We can do the same calculations for $f_{---+}=E_{14,14}$:
\begin{eqnarray}
\psi_{3}(E_{14,14}) &=& 
\tfrac{3}{4} E_{1,1} 
+ \tfrac{1}{4} E_{14,14} 
+ \tfrac{i\sqrt{3}}{4}\,(E_{1,14} - E_{14,1}),\\
\psi_{3}^{2}(E_{14,14}) &=& 
\tfrac{3}{4} E_{1,1} 
+ \tfrac{1}{4} E_{14,14} 
- \tfrac{i\sqrt{3}}{4}\,(E_{1,14} - E_{14,1}),
\end{eqnarray}
and likewise for $f_{+++-}=E_{9,9}$ and $f_{----}=E_{6,6}$. The subspaces
\begin{eqnarray}
    \bb{C}\ell(8)E_{1,1}&\oplus& \bb{C}\ell(8)E_{14,14},\\
    \bb{C}\ell(8)E_{9,9}&\oplus& \bb{C}\ell(8)E_{6,6},
\end{eqnarray}
are therefore each closed under $S_3$. Each of these subspaces is 32 (complex) dimensional. 

One furthermore finds that although the $SU(4)$ generators are not $S_3$-invariant, the $SU(3)$ generators $\Lambda_1,...,\Lambda_8$ are \cite{gourlay2024algebraic}:
\begin{eqnarray}
    \psi_3(\Lambda_i) = \Lambda_i \qquad i=1,\dots,8.
\end{eqnarray}
This is expected since $SU(3)$ corresponds to a maximal subgroup of $\textrm{Aut}(\bb{O})=G_2$ and $S_3$ fixes $G_2$ pointwise.

Starting with the first minimal left ideal $I_1$ and applying $\psi_3$, we can generate three sets of 16 states. These however cannot all be linearly independent, since the $\psi_3$-closed subspace $\bb{C}\ell(8)E_{1,1}\oplus \bb{C}\ell(8)E_{14,14}$ is only 32-dimensional. 

Consider therefore splitting the 16 dimensional spinor $I_1$ into its 8 dimensional even and odd semi-spinors
\begin{eqnarray}
    I_1^+=\rho^+I_1,\qquad I_1^-=\rho^-I_1,
\end{eqnarray}
where $\rho^{\pm}:=\frac{1}{2}(1\pm\omega_8)$. Then,
\begin{eqnarray}
    I_1^+:=\bb{C}\ell^+(8)f_{++++}=(r_0 + r_{ij} a_i^\dagger a_j^\dagger + r_{i4} a_i^\dagger a_4^\dagger+ r_{1234} a_1^\dagger a_2^\dagger a_3^\dagger a_4^\dagger)f_{++++},
\end{eqnarray}
where $i, j \in \{1,2,3\}$, $i < j$ and $r \in \mathbb{C}$. The generated set of 24 states $\{I_1^{+},\psi_3(I_1^+),\psi_3^2(I_1^{+})\}$ are then found to all be linearly independent \cite{gourlay2024algebraic}\footnote{This linear independence calculation was verified in Mathematica; the corresponding notebook is available at:\url{https://gitlab.com/cosurgi/verification-of-arxiv-2601.07857}.}. 


\section{Three generations of fermions with $SU(3)_C\times U(1)_{em}$ gauge symmetry in $\bb{C}\ell(8)$}\label{sec:3genSU(3)}

In an earlier work \cite{gourlay2024algebraic}, we used the above described structure of $S_3$ inside $\bb{C}\ell(8)$ to represent three linearly independent generations of fermions with unbroken $SU(3)_C\times U(1)_{em}$ gauge symmetry. Here, a brief overview of this construction is provided. 



We already established that the $SU(3)$ generators $\Lambda_1,...,\Lambda_8$ are $S_3$-invariant. This is essential to avoid introducing three generations of gluons.

Identify (half of) the first generation states with the first semi-spinor $I_1^{+}$:
\begin{eqnarray}
    I_1^+:=\bb{C}\ell^+(8)f_{++++}=(r_0 + r_{ij} a_i^\dagger a_j^\dagger + r_{i4} a_i^\dagger a_4^\dagger+ r_{1234} a_1^\dagger a_2^\dagger a_3^\dagger a_4^\dagger)f_{++++}.
\end{eqnarray}
These first generation states transform via $SU(3)$ as
\begin{eqnarray}
    I_1^+\cong1 \oplus 3 \oplus \overline{3} \oplus 1,
\end{eqnarray}
as required. The action of $SU(3)$ on $I_1^+$ is taken to be via the commutator $\left[\Lambda_i, I_1^+\right]$, following \cite{chisholm1999gauge}. However it is readily checked that this commutator action reduces to a left-only action, since the term $I_1^+\Lambda_i=0$\footnote{See Section \ref{sec:spinortransform} for a more in-depth discussion.}. That is
\begin{eqnarray}\label{eqn.SU(3)left}
   \left[\Lambda_i, I_1^+\right]=\Lambda_iI_1^+,\quad i=1,...,8. 
\end{eqnarray}

To obtain both particle and anti-particle states we include the odd semi-spinor of the minimal ideal, $I_9^-=I_1^{+}a_4$, constructed on the primitive idempotent $f_{+++-}=E_{9,9}$. These states likewise transform as
\begin{eqnarray}
    I_9^-\cong1 \oplus 3 \oplus \overline{3} \oplus 1,
\end{eqnarray}
where again the commutator action reduces to a left action. The first generation of fermions is therefore represented as
\begin{eqnarray}
    I_1^+\oplus I_9^-&=& \bb{C}\ell^+(8)f_{++++}\oplus \bb{C}\ell^-(8)f_{+++-},\qquad\textrm{First generation.}
\end{eqnarray}
The second and third generations are then obtained via the action of $\psi_3$ on the first generation of states. That is:
\begin{eqnarray}
    &\;&\psi_3(I_1^+\oplus I_9^-),\qquad \textrm{Second generation,}\\
    &\;&\psi_3^2(I_1^+\oplus I_9^-),\qquad \textrm{Third generation.}
\end{eqnarray}

The three sets of Witt bases $\{a_i,a_i^{\dagger}\}$, $\{\psi_3(a_i),\psi_3(a_i^{\dagger})\}$, and $\{\psi^2_3(a_i),\psi^2_3(a_i^{\dagger})\}$ all satisfy the usual fermionic anticommutation relations (\ref{eqn:fermionicalgebra}), and $\psi_3$ transforms primitive idempotents into primitive idempotents. Because the $\psi_3$-invariant spaces $I_1^{+}\oplus\psi_3(I_1^+)\oplus\psi_3^2(I_1^+)$ and $I_9^{-}\oplus\psi_3(I_9^-)\oplus\psi_3^2(I_9^-)$ are linearly independent as subspaces of $\textrm{Mat}(16,\bb{C})$, the 48 states represented in this way are guaranteed to be linearly independent in $\bb{C}\ell(8)$.

Since the $SU(3)$ generators $\Lambda_i$ are fixed by $\psi_3$, the map $\psi_3$ intertwines the $SU(3)$ action. Indeed, for any $X \in I_1^+\oplus I_9^-$ we have
\begin{eqnarray}
    \psi_3([\Lambda_i,X])
    &=& \psi_3(\Lambda_i X - X \Lambda_i) \nonumber\\
    &=& \psi_3(\Lambda_i)\psi_3(X) - \psi_3(X)\psi_3(\Lambda_i) \nonumber\\
    &=& \Lambda_i \psi_3(X) - \psi_3(X)\Lambda_i \nonumber\\
    &=& [\Lambda_i,\psi_3(X)].
\end{eqnarray}
Thus $\psi_3 : I_1^+\oplus I_9^- \to \psi_3(I_1^+\oplus I_9^-)$ is an $SU(3)$-equivariant isomorphism, and the same holds for $\psi_3^2 : I_1^+\oplus I_9^- \to \psi_3^2(I_1^+\oplus I_9^-)$. Since we already know that
\begin{eqnarray}
    I_1^+ \;\cong\; 1 \oplus 3 \oplus \bar{3} \oplus 1,\qquad I_9^- \;\cong\; 1 \oplus 3 \oplus \bar{3} \oplus 1
\end{eqnarray}
as an $SU(3)$ representation, it follows that the second and third generations carry the same decomposition. The three generations are therefore equivalent $SU(3)$ modules.

The $S_3$-invariant electric charge generator is defined as
\begin{eqnarray}
    Q:=\frac{1}{3}\left(Q_1+\psi_3(Q_1)+\psi_3^2(Q_1)\right),
\end{eqnarray}
where
\begin{eqnarray}
    Q_1:=\frac{1}{3}\left(a_1a_1^{\dagger}+a_2a_2^{\dagger}+a_3a_3^{\dagger}-3a_4a_4^{\dagger}\right).
\end{eqnarray}
The generator $Q_1$ is proportional to the $SU(4)$ generator $\Lambda_{15}$ which commutes with $SU(3)$. $I_1^{+}$ then contains the isospin-up states, whereas $I_9^{-}$ contains the isospin-down states. The $Q$ eigenvalues for the first generation of states are listed in Table 1.

\begin{table}[h]\label{table1}
\centering
\begin{tabular}{c  c}
\hline
State &  $Q$ eigenvalue \\
\hline
$f_{++++}$ 
& $0$ \\
$a_i^{\dagger} a_j^{\dagger} f_{++++}$ 
& $\tfrac{1}{3}$ \\
$a_i^{\dagger} a_4^{\dagger} f_{++++}$ 
& $\tfrac{2}{3}$ \\
$a_1^{\dagger} a_2^{\dagger} a_3^{\dagger} a_4^{\dagger} f_{++++}$ 
& $1$ \\
$a_4f_{+++-}$ 
& $-1$ \\
$a_i^{\dagger} a_j^{\dagger} a_4f_{+++-}$ 
& $-\tfrac{2}{3}$ \\
$a_i^{\dagger} f_{+++-}$ 
& $-\tfrac{1}{3}$ \\
$a_1^{\dagger} a_2^{\dagger} a_3^{\dagger} f_{+++-}$ 
& $0$ \\
\hline
\end{tabular}
\caption{Eigenvalues of the adjoint action $[Q,\cdot]$ on first generation states. $i,j\in\{1,2,3\}$, $i<j$.}
\end{table}


\subsection{Spinor transformations and localisation of the gauge action}\label{sec:spinortransform}

In the present framework, we follow \cite{chisholm1999gauge} and take the gauge generators to act on algebraic spinor states through the commutator
\begin{eqnarray}
[\Lambda_a,X] , \qquad X \in \mathbb{C}\ell(8).
\end{eqnarray}
This commutator in general mixes left and right multiplication. Whereas left multiplication leads to a transformation within a given minimal left ideal, right multiplication leads to transformations between different minimal ideals.

For the algebraic states of interest, however, this commutator localises to a pure left action on precisely four minimal left ideals,
\begin{eqnarray}
I_1&=&\bb{C}\ell(8)f_{++++},\qquad I_{14}=\bb{C}\ell(8)f_{---+},\\ 
I_9&=&\bb{C}\ell(8)f_{+++-},\qquad I_6=\bb{C}\ell(8)f_{----}, 
\end{eqnarray}
corresponding to the four spinors used in our construction of three fermion generation with $SU(3)_C\times U(1)_{em}$ gauge symmetry. On these ideals the action of the $SU(3)$ generators is implemented entirely by left multiplication, so the fermionic states transform as genuine left-modules under the colour gauge algebra.

More explicitly, for any generator $\Lambda_a$ with $a \in \{1,\dots,8\}$ one has
\begin{equation}
[\Lambda_a,X]=\Lambda_a X \qquad \text{for all }  X\in I_1\oplus I_{14}\oplus I_6\oplus I_9.
\end{equation}

On minimal ideals other than $I_1,I_6,I_9,I_{14}$, the commutator is not purely
left-acting in general. For the Cartan generators $\Lambda_3$ and $\Lambda_8$, an ideal-dependent left-acting
representative exists on every minimal left ideal. Indeed, since $\Lambda_3$ and $\Lambda_8$
are diagonal in the basis of eqn.~(\ref{Cl(8)basis}), for each minimal left ideal
$I_j=\bb{C}\ell(8)E_{j,j}$ there exist scalars $\lambda_r^{(j)}$ such that
\begin{eqnarray}
[\Lambda_r,X] = \bigl(\Lambda_r-\lambda_r^{(j)}I\bigr)X=\Lambda^{[j]}_rX,
\qquad r\in\{3,8\},\; X\in I_j.
\label{eq:CartanLocal}
\end{eqnarray}
These replacements $\Lambda^{[j]}_3$ and $\Lambda^{[j]}_8$ differ from the original $\Lambda_3,\Lambda_8$ only by elements of the commutant of $I_j$, and therefore do not modify the physical
representation of the colour algebra. In particular, the redefined Cartan generators close the same $\mathfrak{su}(3)$ Lie algebra, and introduce no new gauge degrees of freedom.

However, this property is special to the Cartan generators. For every minimal left ideal
other than $I_1,I_6,I_9,I_{14}$, at least one non-Cartan generator
$\Lambda_a$ with $a\in\{1,2,4,5,6,7\}$ has unavoidable off-diagonal support into other columns (minimal left ideals), and the commutator fails to restrict to a single left multiplier.

For the $U(1)_{em}$ generator $Q$, the commutator $[Q,X]$ does not a priori reduce to a purely left action.
However, $Q$ is diagonal in the basis in eqn. (\ref{Cl(8)basis}), and for all algebraic states $X$ belonging to the minimal left ideals used in the present construction one finds
\begin{equation}
XQ =
\begin{cases}
-\frac{1}{2}X, & X \in I_1 \oplus I_{14},\\
\phantom{-}\frac{1}{2}X, & X \in I_6 \oplus I_9.
\end{cases}
\end{equation}

As a consequence, the right action of $Q$ reduces to a constant on each of the two
$\psi_3$-closed subspaces $I_1\oplus I_{14}$ and $I_6\oplus I_9$. On the physically relevant minimal left ideals, the commutator may therefore be rewritten as
\begin{equation}
[Q,X] =
\begin{cases}
\left(Q+\frac{1}{2}I\right)X, & X \in I_1 \oplus I_{14},\\
\left(Q-\frac{1}{2}I\right)X, & X \in I_6 \oplus I_9.
\end{cases}
\end{equation}
This defines an equivalent purely left-acting generator. Since the identity element $\bb{I}$ is central in $\bb{C}\ell(8)$, this shift does not modify the $U(1)_{em}$ algebra or the electric charge assignments extracted from the commutator action. 

Thus, on the physically relevant minimal ideals, both the colour and electromagnetic gauge symmetry acts entirely by left multiplication, and introduces no mixing between distinct minimal left ideals.

Finally, we note that commutator action of the $SU(2)_L$ symmetry to be introduced in Section \ref{sec:chiralSU(2)} likewise reduces to a one-sided action. In that case however, we find that the commutator action reduces to a right action. Consequently, the $SU(2)_L$ action does not preserve individual minimal left ideals, but instead induces transformations between distinct ideals.


\section{Enlarging $\bb{C}\ell(8)$ to $\bb{C}\ell(10)$}

In order to include the weak sector, we need to enlarge the algebra from $\bb{C}\ell(8)$ to $\bb{C}\ell(10)$\footnote{The algebra $\bb{C}\ell(8)$ corresponds to the algebra of endomorphisms of $\bb{C}\otimes\bb{S}$. Likewise, $\bb{C}\ell(10)$ an be considered as the left (only) endomorphism algebra of $\bb{C}\otimes\bb{H}\otimes\bb{S}$.}. We achieve this using the graded tensor product realization $\bb{C}\ell(10)\cong\bb{C}\ell(2)\;\hat{\otimes}\;\bb{C}\ell(8)$. Consider the unital injective homomorphism
\begin{eqnarray}
    \iota:\bb{C}\ell(8)&\hookrightarrow& \bb{C}\ell(10),\\
    \iota(e_i)&\mapsto& \bb{I}_2\otimes e_i,\quad i=1,...,8, 
\end{eqnarray}
and define the remaining $\bb{C}\ell(10)$ basis units as
\begin{eqnarray}
    e_9:=-i\bar{\sigma}_1\otimes \omega_8,\qquad e_{10}:=i\bar{\sigma}_2\otimes \omega_8,
\end{eqnarray}
where $\bar{\sigma}_i$ are Pauli matrices, and the bar is used to avoid confusion with the Pauli matrices used to define our $\bb{C}\ell(8)$ basis in eqn. (\ref{Cl(8)basis}). From now on, unless confusion is likely to arise, we will simply write $e_i$ instead of $\iota(e_i)$.

Note that 
\begin{eqnarray}\label{eq:blockdecomposition}
    \iota(\bb{C}\ell(8))=\bb{I}_2\otimes\bb{C}\ell(8)=\begin{pmatrix}\bb{C}\ell(8) & 0\\ 0 & \bb{C}\ell(8)\end{pmatrix},
\end{eqnarray}
so that $\bb{C}\ell(8)$ once embedded into $\bb{C}\ell(10)$ takes on a block-diagonal form.

We can use the same injective homomorphism $\iota$ to define the $\bb{C}\ell(10)$ Witt basis
\begin{eqnarray}
        \iota(a_j) &=& \frac{1}{2}(-\iota(e_j) + i\iota(e_{j+4})),\quad a_5:=\frac{1}{2}(-e_9+ie_{10}), \\
        \iota(a_j)^\dagger &=& \frac{1}{2}(\iota(e_j) + i\iota(e_{j+4})),\quad a_5^{\dagger}:=\frac{1}{2}(e_9+ie_{10}).
\end{eqnarray}
For simplicity we will from now on simply write (for example) $a_i$ instead of $\iota(a_i)$ unless confusion is likely to arise. In particular one finds that
\begin{eqnarray}
    a_5=\begin{pmatrix}
        0 & i\omega_8\\
        0 &0
    \end{pmatrix},\qquad a_5^{\dagger}=\begin{pmatrix}
        0 & 0\\
        -i\omega_8 &0
        \end{pmatrix}.
\end{eqnarray}
Next, define the simple idempotents
\begin{eqnarray}
    \pi^{(+)}_5:=a_5a_5^{\dagger}=\begin{pmatrix}
        \bb{I}_{16} & 0\\
        0& 0
    \end{pmatrix},\qquad \pi^{(-)}_5:=a_5^{\dagger}a_5=\begin{pmatrix}
         0 & 0\\
        0& \bb{I}_{16}
    \end{pmatrix}.
\end{eqnarray}
Subsequently the 32 primitive idempotents of $\bb{C}\ell(10)$ are given by
\begin{eqnarray}
    f_{\varepsilon_1\varepsilon_2\varepsilon_3\varepsilon_4\varepsilon_5}
\;=\;
\pi_1^{(\varepsilon_1)}\pi_2^{(\varepsilon_2)}\pi_3^{(\varepsilon_3)}\pi_4^{(\varepsilon_4)}\pi_5^{(\varepsilon_5)},
\qquad \varepsilon_j\in\{+,-\}.
\end{eqnarray}
They are mutually orthogonal and complete:
\begin{equation}
f_{\varepsilon} f_{\varepsilon'} = \delta_{\varepsilon_1,\varepsilon'_1}\delta_{\varepsilon_2,\varepsilon'_2}\delta_{\varepsilon_3,\varepsilon'_3}\delta_{\varepsilon_4,\varepsilon'_4}\delta_{\varepsilon_5,\varepsilon'_5} f_{\varepsilon},
\qquad
\sum_{\varepsilon\in\{+,-\}^5} f_{\varepsilon} = \bb{I}_{32}, \qquad f_{\varepsilon}=f_{\varepsilon_1\varepsilon_2\varepsilon_3\varepsilon_4\varepsilon_5}
\end{equation}
and correspond to rank-1 diagonal projectors equal to matrix units $E_{i,i}$ where $i=1,...,32$.

It is readily checked that
\begin{eqnarray}
    f_{+++++}:=f_{++++}\pi_5^{(+)}=E_{1,1},\qquad f_{++++-}:=f_{++++}\pi_5^{(-)}=E_{17,17}.
\end{eqnarray}
More generally we find that
\begin{eqnarray}
    \iota(E_{i,i})\pi_5^{(+)}=E_{i,i},\qquad \iota(E_{i,i})\pi_5^{(-)}=E_{i+16,i+16},\qquad i=1,...,16.
\end{eqnarray}
We can therefore write the first minimal left ideal of $\bb{C}\ell(10)$ as
\begin{eqnarray}
    I_1:=\bb{C}\ell(10)f_{+++++}&=&\bb{C}\ell(8)f_{+++++}\oplus \bb{C}\ell(8)a_5^{\dagger}f_{+++++},\\&=&\begin{pmatrix}\bb{C}\ell(8)f_{++++} & 0\\ 0 & 0\end{pmatrix}\oplus \begin{pmatrix}0 & 0\\ \bb{C}\ell(8)a_5^{\dagger}f_{++++} & 0\end{pmatrix}.
\end{eqnarray}

\subsection{Embedding $S_3$ into $\bb{C}\ell(10)$}

The $S_3$ generators $\psi_3$ and $\epsilon$ act on the $\bb{C}\ell(8)$ basis $\{e_i\}$, whereas their action on the $\bb{C}\ell(2)$ algebra generated by $\{\bar{\sigma}_a\}$, $a=1,2,3$ is trivial
\begin{eqnarray}
    \psi_3(\bar{\sigma}_a)=\bar{\sigma}_a,\qquad \epsilon(\bar{\sigma}_a)=\bar{\sigma}_a,\qquad a=1,2,3.
\end{eqnarray}
Nonetheless, the action of $S_3$ on $e_9$ and $e_{10}$ is nontrivial, since both include a factor of $\omega_8$, the $\bb{C}\ell(8)$ pseudoscalar which is itself not invariant under $\psi_3$.
\begin{eqnarray}
    \psi_3(e_9)&=&\psi_3(-i\bar{\sigma}_1\otimes \omega_8)=-i\bar{\sigma}_1\otimes \psi_3(\omega_8),\\
    \psi_3(e_{10})&=&\psi_3(i\bar{\sigma}_2\otimes \omega_8)=i\bar{\sigma}_2\otimes \psi_3(\omega_8).
\end{eqnarray}
On the other hand $\epsilon(\omega_8)=\omega_8$, and so
\begin{eqnarray}
    \epsilon(a_5)=a_5,\qquad \epsilon(a_5^{\dagger})=a_5^{\dagger}.
\end{eqnarray}

It can subsequently be verified that the three Witt bases $\{a_i,a_i^{\dagger}\}$, $\{\psi_3(a_i),\psi_3(a_i^{\dagger})\}$, and $\{\psi^2_3(a_i),\psi^2_3(a_i^{\dagger})\}$ (now with $i=1,...,5$) all satisfy the usual fermionic anticommutation relations, and that $\psi_3$ transforms primitive idempotents into primitive idempotents.

The $S_3$ action is block diagonal with respect to the $\bb{C}\ell(2)\hat{\otimes}\bb{C}\ell(8)$ decomposition in eqn. (\ref{eq:blockdecomposition}). In other words
\begin{eqnarray}
    \iota:\psi_3&\hookrightarrow& \psi_3\oplus \psi_3,\\
    \iota:\epsilon&\hookrightarrow&\epsilon\oplus\epsilon.
\end{eqnarray}
In $\bb{C}\ell(8)$ it was established that the subspaces
\begin{eqnarray}
    \bb{C}\ell(8)f_{++++}\oplus \bb{C}\ell(8)f_{---+},\\
     \bb{C}\ell(8)f_{+++-}\oplus \bb{C}\ell(8)f_{----},
\end{eqnarray}
are each closed under $S_3$. When we subsequently pass to $\bb{C}\ell(10)$ via the graded tensor product, we obtain four copies of this structure. Explicitly, the (64 complex dimensional) closed subspaces are
\begin{eqnarray}
    \bb{C}\ell(10)f_{+++++}\oplus \bb{C}\ell(10)f_{---++}&=&\bb{C}\ell(10)E_{1,1}\oplus \bb{C}\ell(10)E_{14,14},\\
    \bb{C}\ell(10)f_{++++-}\oplus \bb{C}\ell(10)f_{---+-}&=&\bb{C}\ell(10)E_{17,17}\oplus \bb{C}\ell(10)E_{30,30},\\
    \bb{C}\ell(10)f_{+++-+}\oplus \bb{C}\ell(10)f_{----+}&=&\bb{C}\ell(10)E_{9,9}\oplus \bb{C}\ell(10)E_{6,6},\\
    \bb{C}\ell(10)f_{+++--}\oplus \bb{C}\ell(10)f_{-----}&=&\bb{C}\ell(10)E_{25,25}\oplus \bb{C}\ell(10)E_{22,22}.
\end{eqnarray}

\section{Three generations of fermion states in $\bb{C}\ell(10)$}

We are now in a position to extend the $\bb{C}\ell(8)$ construction of Section \ref{sec:3genSU(3)} to $\bb{C}\ell(10)$ as a way to incorporate the $SU(2)_L\times U(1)_Y$ gauge sector of the SM. The goal is to 
 embed the SM gauge symmetries into $\bb{C}\ell(10)$ in such a way that
\begin{enumerate}
    \item The three generations of physical states are linearly independent,
    \vspace{-10pt}
    \item The Lie algebra generators of the gauge symmetries are invariant under $S_3$,
    \vspace{-10pt}
    \item The commutator action of gauge symmetries on fermion states reduce to a one sided action (subject to the discussion in Section \ref{sec:spinortransform}),
    \vspace{-10pt}
    \item The gauge groups commute.
\end{enumerate}

\subsection{The colour symmetry $SU(3)_C$}
The $SU(3)$ generators used in the $\bb{C}\ell(8)$ model embed into $\bb{C}\ell(10)$ in a straighforward manner
\begin{eqnarray}
    \iota:\Lambda_i\in \bb{C}\ell(8)\hookrightarrow \bb{I}_2\otimes \Lambda_i\in \bb{C}\ell(10),\qquad i=1,...,8.
\end{eqnarray}
These generators therefore remain invariant under $S_3$ and $SU(3)$ now acts identically on the two $\bb{C}\ell(8)$ blocks in eqn. (\ref{eq:blockdecomposition}).

We can then embed the three linearly independent generations of fermions from $\bb{C}\ell(8)$ into $\bb{C}\ell(10)$. There is not a unique way to do this, but one option is to embed these states in the upper $\bb{C}\ell(8)$ block of (\ref{eq:blockdecomposition}) via right multiplication by $\pi_5^{(+)}$:
\begin{eqnarray}
    \iota(I_1^+)\pi_5^{(+)}\oplus \iota(I_{9}^-)\pi_5^{(+)},\qquad \textrm{First generation.}
\end{eqnarray}
The remaining two generations are then obtained via $\psi_3$. This gives the required 48 linearly independent states that transform as three generations under $SU(3)$.

\subsection{The chiral weak symmetry $SU(2)_L$}\label{sec:chiralSU(2)}
In order to include the weak symmetry the number of states needs to be doubled to ensure each fermions comes in both left- and right-handed copies. To ensure $SU(2)_L$ only acts non-trivially on the left-handed states, an ($S_3$-invariant) projector will be introduced. The role of this projector will be to identify the minimal left ideals which contain $SU(2)_L$ doublet states. To ensure that the $SU(2)_L$ generators introduced in the next subsection are $S_3$ invariant, the projector itself needs to be $S_3$ invariant, and commute with the $SU(3)$ generators.

\subsubsection{An $S_3$ invariant projector}
In section \ref{sec:S3} the action of $\psi_3$ on the $\bb{C}\ell(8)$ primitive idempotent $f_{++++}=E_{1,1}$ was determined. In particular the rank-2 projector $f_{++++}+f_{---+}=E_{1,1}+E_{14,14}$ was found to be $S_3$-invariant. Since $a_5^{\dagger}a_5$ (and $a_5a_5^{\dagger}$) are likewise $S_3$-invariant, the action of $\psi_3$ (and $\epsilon$) on $f_{++++}$ carries over to $f_{+++++}$ in $\bb{C}\ell(10)$.  Subsequently, an $S_3$-invariant rank-4 projector can be defined in $\bb{C}\ell(10)$:
\begin{eqnarray}
    P&:=&\bb{I}_2\otimes (f_{++++}+f_{---+})=f_{+++++}+f_{---++}+f_{++++-}+f_{---+-},\\
    &=&E_{1,1}+E_{14,14}+E_{17,17}+E_{30,30}\in\bb{C}\ell(10).
\end{eqnarray}
All four primitive idempotents that make up $P$ transform as $SU(3)$ singlets, guaranteeing that $P$ commutes with $SU(3)$ as required. It is furthermore readily verified that
\begin{eqnarray}
    P\omega_8=\omega_8P,\quad Pe_9=e_9P,\quad Pe_{10}=e_{10}P, 
\end{eqnarray}
from which it follows that $P$ commutes with both $a_5$ and $a_5^{\dagger}$. It is important to note that
\begin{eqnarray}
    Pf_{\varepsilon_1\varepsilon_2\varepsilon_3\varepsilon_4\varepsilon_5}=f_{\varepsilon_1\varepsilon_2\varepsilon_3\varepsilon_4\varepsilon_5}P=f_{\varepsilon_1\varepsilon_2\varepsilon_3\varepsilon_4\varepsilon_5}
\end{eqnarray}
if $\varepsilon_1\varepsilon_2\varepsilon_3\varepsilon_4\varepsilon_5\in\small{\{+++++,---++,++++-,---+-\}}$, and is equal to zero otherwise.


\subsubsection{$S_3$ invariant $SU(2)_L$ generators}

The $SU(2)_L$ generators can now be defined as
\begin{eqnarray}
    T_1:&=&\frac{1}{2}(-ia_5+ia_5^{\dagger})\omega_8P=\frac{i}{2}e_9\omega_8P,\\
    T_2:&=&\frac{1}{2}(a_5+a_5^{\dagger})\omega_8P=\frac{i}{2}e_{10}\omega_8P,\\
    T_3:&=&\frac{1}{2}(a_5^{\dagger}a_5-a_5a_5^{\dagger})P=\frac{i}{2}e_9e_{10}P,
\end{eqnarray}
from which it follows that
\begin{eqnarray}
    T_{+}:=T_1+ iT_2=ia_5^{\dagger}\omega_8P,\qquad T_{-}:=T_1-iT_2=-ia_5\omega_8P.
\end{eqnarray}
These generators commute with the $SU(3)_C$ generators.

We already established that $P$ is an $S_3$-invariant projector. The $S_3$-invariance of $e_9\omega_8$ and $e_{10}\omega_8$ follows immediately from the definition of $e_9$ and $e_{10}$:
\begin{eqnarray}
   (-i\bar{\sigma}_1\otimes \omega_8)(\bb{I}_2\otimes\omega_8)&=&-i\bar{\sigma}_1\otimes\bb{I}_{16},\\
   (\bar{\sigma}_2\otimes \omega_8)(\bb{I}_2\otimes\omega_8)&=&\bar{\sigma}_2\otimes\bb{I}_{16}.
\end{eqnarray}
The $SU(2)_L$ generators are therefore invariant under $S_3$, as required in order to avoid introducing three generations of weak gauge bosons.


\subsubsection{Weak doublet states}
We are now in a position to determine our physical states. Focus first on the eight isospin-up states of the first generation. We need these states to satisfy the following properties:
\begin{enumerate}
    \item They must transform as $1\oplus3\oplus\bar{3}\oplus1$ via $SU(3)$,
    \vspace{-10pt}
    \item The 24 states obtained from applying $\psi_3$ must be linearly independent,
    \vspace{-10pt}
    \item Either the left or right action of the projector $P$ onto the states must give zero in order to ensure the the commutator action of $SU(2)_L$ on the states reduces to a one-sided action,
    \vspace{-10pt}
    \item The states must give the correct electric charges.
\end{enumerate}

In $\bb{C}\ell(8)$, the states $I_1^+$ were shown to transform as $1\oplus3\oplus\bar{3}\oplus1$ via $SU(3)$, and the 24 states $\{I_1^+,\psi_3(I_1^+),\psi_3^2(I_1^+)\}$ generated via $\psi_3$ were shown to be linearly independent. The most straightforward choice is therefore to consider the states $\iota(I_1^{+})\pi_5^+\in\bb{C}\ell(10)$ as the isospin-up states of the first generation. This however does not work because 
\begin{eqnarray}
    \iota(I_1^{+})\pi_5^+P=\iota(I_1^{+})\pi_5^+,\qquad P\iota(I_1^{+})\pi_5^+\propto f_{+++++}.
\end{eqnarray}
The commutator action of $SU(2)_L$ on these states therefore does not reduce to a one-sided action. The only offending term is the $SU(3)$ singlet state $f_{++++}$. We might therefore find a replacement singlet state in $I_1\in\bb{C}\ell(10)$. Let us therefore replace the state $f_{++++}$ with $a_4^{\dagger}a_5^{\dagger}f_{++++}$.

Explicitly, we choose our eight isospin-up states (of the first generation) as follows:
\begin{eqnarray}
    a_1^{\dagger}a_2^{\dagger}a_3^{\dagger}a_4^{\dagger}f_{+++++}&,&\quad (SU(3) \;\textrm{singlet}),\\
    a_i^{\dagger}a_4^{\dagger}f_{+++++}&,&\quad (SU(3)\;\textrm{triplet}),\\
    a_i^{\dagger}a_j^{\dagger}f_{+++++}&,&\quad (SU(3) \;\textrm{anti-triplet}),\\
    a_4^{\dagger}a_5^{\dagger}f_{+++++}&,&\quad(SU(3)\;\textrm{singlet}),
\end{eqnarray}
with $i,j\in\{1,2,3\}$ and $i<j$. Note that all these states have even grade and so live in the even semi-spinor $I_1^+$ of $\bb{C}\ell(10)$.

Define
\begin{eqnarray}
    V_1^+:=(c_{1234}a_1^{\dagger}a_2^{\dagger}a_3^{\dagger}a_4^{\dagger}+c_{i4}a_i^{\dagger}a_4^{\dagger}+c_{ij}a_i^{\dagger}a_j^{\dagger}+c_{45}a_4^{\dagger}a_5^{\dagger})f_{+++++},\quad i,j=1,2,3,\quad i<j
\end{eqnarray}
where $c_{1234},c_{i4},c_{ij},c_{45}\in\bb{C}$. Via the action of $\psi_3$, the resulting 24 states are linearly independent and live entirely in the closed subspace spanned by the two minimal ideals built on the primitive idempotents $f_{+++++}=E_{1,1}$ and $f_{---++}=E_{14,14}$. This ensures three linearly independent generations of isospin-up states. 

Define therefore the 24 isospin-up states as
\begin{eqnarray}
    V^+:=V_1^+\oplus \psi_3(V_1^+)\oplus\psi_3^2(V_1^+)\subset \bb{C}\ell(10)f_{+++++}\oplus\bb{C}\ell(10)f_{---++}.
\end{eqnarray}

The corresponding isospin down states are obtained simply by multiplying the isospin-up states on the right by $a_5$.
\begin{eqnarray}
    a_1^{\dagger}a_2^{\dagger}a_3^{\dagger}a_4^{\dagger}f_{+++++}a_5&=&a_1^{\dagger}a_2^{\dagger}a_3^{\dagger}a_4^{\dagger}a_5f_{++++-},\quad (SU(3) \;\textrm{singlet}),\\
    a_i^{\dagger}a_4^{\dagger}f_{+++++}a_5&=&a_i^{\dagger}a_4^{\dagger}a_5f_{++++-},\quad (SU(3)\;\textrm{triplet}),\\
    a_i^{\dagger}a_j^{\dagger}f_{+++++}a_5&=&a_i^{\dagger}a_j^{\dagger}a_5f_{++++-},\quad (SU(3) \;\textrm{anti-triplet}),\\
    a_4^{\dagger}a_5^{\dagger}f_{+++++}a_5&=&a_4^{\dagger}f_{++++-},\quad(SU(3)\;\textrm{singlet}),
\end{eqnarray}
where we have used $f_{+++++}a_5=f_{++++}a_5a_5^{\dagger}a_5=a_5f_{++++}a_5^{\dagger}a_5=a_5f_{++++-}$\footnote{More generally one finds that 
\begin{eqnarray}
    f_{+++++}a_5^{\dagger}&=&0,\quad f_{+++++}a_5=a_5f_{++++-},\\
    f_{++++-}a_5&=&0,\quad f_{++++-}a_5^{\dagger}=a_5^{\dagger}f_{+++++},\\
    f_{----+}a_5^{\dagger}&=&0,\quad f_{----+}a_5=a_5f_{-----},\\
    f_{-----}a_5&=&0,\quad f_{-----}a_5^{\dagger}=a_5^{\dagger}f_{----+}.
\end{eqnarray}.}. 

That is, right multiplying states in the minimal left ideal $I_1$ by $a_5$ gives states in the minimal left ideal $I_{17}$ (built on the primitive idempotent $f_{++++-}$). Likewise, right multiplying states in $I_{17}$ by $a_5^{\dagger}$ gives back states in $I_1$.
\begin{eqnarray}
    \bb{C}\ell(10)\underbrace{f_{+++++}}_{\displaystyle E_{1,1}}a_5&=&\bb{C}\ell(10)\underbrace{f_{++++-}}_{\displaystyle E_{17,17}},\quad \bb{C}\ell(10)\underbrace{f_{++++-}}_{\displaystyle E_{17,17}}a_5^{\dagger}=\bb{C}\ell(10)\underbrace{f_{+++++}}_{\displaystyle E_{1,1}}.
\end{eqnarray}

Define
\begin{eqnarray}
    V_1^-:=V_1^+a_5=(c_{12345}a_1^{\dagger}a_2^{\dagger}a_3^{\dagger}a_4^{\dagger}a_5+c_{i45}a_i^{\dagger}a_4^{\dagger}a_5+c_{ij5}a_i^{\dagger}a_j^{\dagger}a_5+c_{4}a_4^{\dagger})f_{++++-},
\end{eqnarray}
where $c_{12345},c_{i45},c_{123},c_{ij}\in\bb{C}$. Applying $\psi_3$ to the isospin-down states results in a total of 24 linearly independent algebraic states that live entirely in the closed subspace spanned by the the two minimal left ideals built on the primitive idempotents $f_{++++-}=E_{17,17}$ and $f_{---+-}=E_{30,30}$. This ensures three linearly independent generations of isospin-down states, that are also disjoint from the isospin-up states. We therefore define the set of isospin-down states as:
\begin{eqnarray}
    V^-:=V_1^-\oplus \psi_3(V_1^-)\oplus\psi_3^2(V_1^-)\subset \bb{C}\ell(10)f_{++++-}\oplus\bb{C}\ell(10)f_{---+-}.
\end{eqnarray}

The $SU(2)_L$ doublet states of the first generation are then
\begin{eqnarray}
    \begin{pmatrix}
        a_1^{\dagger}a_2^{\dagger}a_3^{\dagger}a_4^{\dagger}f\\
        a_1^{\dagger}a_2^{\dagger}a_3^{\dagger}a_4^{\dagger}fa_5
    \end{pmatrix},\; 
    \begin{pmatrix}
        a_i^{\dagger}a_4^{\dagger}f\\
       a_i^{\dagger}a_4^{\dagger}fa_5
    \end{pmatrix},\;
     \begin{pmatrix}
        a_i^{\dagger}a_j^{\dagger}f\\
       a_i^{\dagger}a_j^{\dagger}fa_5
    \end{pmatrix},\;
    \begin{pmatrix}
        a_4^{\dagger}a_5^{\dagger}f\\
       a_4^{\dagger}a_5^{\dagger}fa_5
    \end{pmatrix},\quad SU(2) \;\textrm{doublets}
\end{eqnarray}
where $f=f_{+++++}$. The second and third generation doublets are obtained by applying $\psi_3$.

For both the isospin-up and isospin-down states, the commutator action of $SU(2)_L$ reduces to a right action
\begin{eqnarray}
    \left[T_i,V^+\right]=-V^+T_i,\qquad \left[T_i,V^-\right]=-V^-T_i,\quad i=1,2,3.
\end{eqnarray}

\subsubsection{Weak singlet states}
To include the weak singlet states, we must identify suitable minimal ideals $I_X$, $X=1,...,32$, such that
\begin{eqnarray}
    I_XP=PI_X=0.
\end{eqnarray}
This condition is satisfied provided that $X\neq 1,14,17,30$. However, the action of $SU(3)$ only reduces to a left-sided action (see eqn. (\ref{eqn.SU(3)left})) for $X=1,6,9,14,17,22,25,30$, leaving us with only the choices $X=6,9,22,25$.

Define the singlet states as 
\begin{eqnarray}
    U_1^+:=V_1^+a_4a_5,\qquad U_1^-:=V_1^+a_4,
\end{eqnarray}
where we note that
\begin{eqnarray}
    f_{+++++}a_4a_5&=&a_4a_5f_{+++--}=a_4a_5E_{25,25},\\
    f_{+++++}a_4&=&a_4f_{+++-+}=a_4E_{9,9}.
\end{eqnarray}
By applying $\psi_3$ we obtain a total of 48 linearly independent singlet states, which are disjoint from the doublet states. Therefore all 96 states are linearly independent. 

Finally therefore, define the three generations of fermion states as follows
\begin{eqnarray}
   S_1:&=&V_1^+\oplus V_1^+a_5\oplus V_1^+a_4\oplus V_1^+a_4a_5,\quad \textrm{First generation,}\\
   S_2:&=&\psi_3(S_1),\quad \textrm{Second generation,}\\
   S_3:&=&\psi_3^2(S_1),\quad \textrm{Third generation.}
\end{eqnarray}
The order-three generator $\psi_3$ then cyclically permutes the three generations of states
\begin{eqnarray}
    S_1 \xrightarrow{\psi_3} S_2 \xrightarrow{\psi_3} S_3 \xrightarrow{\psi_3}S_1.
\end{eqnarray}

\subsection{Electric charge $U(1)_{em}$ and hypercharge $U(1)_Y$}
Using the electric charge generator $Q$ from Section \ref{sec:3genSU(3)} assigns the same electric charges to the isospin-up and isospin-down states, because $Q$ commutes with $a_5$. We therefore define the electric charge generator in $\bb{C}\ell(10)$ as
\begin{eqnarray}
    Q':&=&Q+(2P-\bb{I})a_5^{\dagger}a_5\\
    &=&\frac{1}{3}\left(Q_1+\psi_3(Q_1)+\psi_3^2(Q_1)\right)+(2P-\bb{I})a_5^{\dagger}a_5, 
\end{eqnarray}
where $Q_1=\frac{1}{3}\left(a_1a_1^{\dagger}+a_2a_2^{\dagger}+a_3a_3^{\dagger}-3a_4a_4^{\dagger}\right)$. Note that $Q'$ remains $S_3$-invariant.

Subsequently, define the hypercharge generator $Y$ as
\begin{eqnarray}
    Y:=2(Q'-T_3).
\end{eqnarray}

The eigenvalues for the first generation of algebraic states, which for definiteness we identify these with the first generation of SM fermions are listed in Table 2 .

\section{Discussion}

In this work we have presented an explicit algebraic framework in which three fermion generations arise intrinsically within a single Clifford algebra and transform under the full SM gauge group. Fermionic states are realised as minimal left ideals of the complex Clifford algebra $\bb{C}\ell(10)$, while the three-generation structure emerges from an embedded discrete $S_3$ symmetry acting on the space of algebraic spinors. A central feature of the construction is that the SM gauge generators are required to be $S_3$-invariant, ensuring that the gauge sector remains unique and is not replicated across generations. In this way, three linearly independent but gauge-equivalent fermion generations are obtained without introducing additional gauge degrees of freedom or ad hoc assumptions.

A notable aspect of the present framework is that it simultaneously realises three fermion generations, the full SM gauge group $SU(3)_C\times SU(2)_L\times U(1)_Y$, and a non-replicated gauge sector within a single Clifford algebra. The generations arise intrinsically from the internal $S_3$ symmetry, while the requirement of $S_3$-invariance of the gauge generators enforces the uniqueness of the gauge sector. Unlike conventional discrete family-symmetry models, where gauge fields are typically assumed to be family blind by construction, the present framework implements this property as a direct algebraic consequence.

Although the construction is formulated within the relatively large algebra $\bb{C}\ell(10)$, the resulting fermionic spectrum is in fact highly constrained. The choice of $\bb{C}\ell(10)$ is minimal in the sense that it is the smallest Clifford algebra in which the colour, electroweak, and family structures can be realised simultaneously within the present framework. In particular, compatibility between the adjoint action of the $SU(3)_C$ generators and their interpretation as left actions severely restricts the available minimal left ideals. Within these ideals, further nontrivial constraints arise: each admissible ideal must support the correct $SU(3)_C$ representations, decomposing as $1 \oplus 3 \oplus \bar{3} \oplus 1$, and the action of the $S_3$ symmetry must generate three generations whose states are linearly independent. Imposing these conditions leaves very little freedom in the choice of physical states, highlighting that the construction is highly constrained rather than arbitrary.

A further structural feature of the construction concerns the realisation of the weak $SU(2)_L$ symmetry. In contrast to the colour sector, whose commutator action can be identified with a left action on the relevant minimal ideals, the $SU(2)_L$ commutator action naturally reduces to a right action. This distinction is not imposed by hand but follows from the presence of an $S_3$-invariant projector contained in the weak subalgebra, which in the present construction has rank four and selects precisely the minimal ideals supporting the electroweak doublet states. By contrast, a left action of $SU(2)_L$ would require an $S_3$-invariant rank-eight projector capable of selecting a full $1 \oplus 3 \oplus \bar{3} \oplus 1$ decomposition within each ideal. No such projector exists in the present framework. The appearance of $SU(2)_L$ as a right action is therefore a structural consequence of the algebraic constraints, rather than an arbitrary modelling choice.

It is also worth contrasting the present framework with conventional $\mathrm{Spin}(10)$ grand unified theories. Although the construction is formulated within the Clifford algebra $\bb{C}\ell(10)$, the Standard Model gauge symmetries do not combine into a single simple gauge group such as $\mathrm{Spin}(10)$, nor is such a unification required here. Instead, the colour and electroweak gauge sectors are realised as distinguished subalgebras whose actions are constrained by compatibility with the $S_3$ family symmetry and the structure of the minimal ideals. The emphasis of the present work is therefore not on gauge unification in the traditional GUT sense, but on identifying an algebraic setting in which gauge structure and fermion generations emerge together.

Beyond its structural features, the present framework suggests several phenomenological directions. One immediate aspect is the presence of an intrinsic $S_3$ family symmetry, which has been widely studied in flavour physics as a means of constraining fermion masses and mixing patterns \cite{gonzalez2013s3, vien2014neutrino, babu2024fermion,hernandez2012lepton,cogollo2016two}. In most existing models, however, $S_3$ is introduced in an ad hoc manner as an external flavour symmetry acting on otherwise conventional gauge representations. In contrast, in the present construction the $S_3$ symmetry arises internally as an automorphism of the algebra and acts directly on the space of algebraic spinors. As a result, the family symmetry is tightly linked to the underlying algebraic structure rather than imposed phenomenologically, which may lead to distinctive constraints on inter-generational couplings once Yukawa interactions and symmetry-breaking effects are incorporated.

A second, closely related feature is the appearance of generation-dependent charge operators. Although the physical electric charge and hypercharge operators are required to be $S_3$-invariant, they are constructed as symmetric combinations of three generation-specific operators (such as $Q_1$). This structure is reminiscent of so-called tri-hypercharge models, in which different generations couple to distinct abelian charges that combine into the observed hypercharge \cite{fernandez2024minimal, navarro2023tri, allanach2019anomaly,martinez2018b}. Such models have been explored previously in phenomenological contexts, but typically without an accompanying discrete family symmetry. In the present framework, the $S_3$ family symmetry and the tri-hypercharge structure arise simultaneously and are closely related: the same symmetry that permutes the fermion generations also enforces the invariance of the physical gauge charges. This interplay between discrete family symmetry and generation-dependent abelian charges appears to be a novel feature of the construction and may have implications for flavour structure and anomaly constraints in extended versions of the model. The phenomenological consequences of these features, including their implications for flavour structure and charge assignments, are currently under investigation and will be presented in future work.

One additional direction for future work is to clarify the relationship between the $S_3$ symmetry appearing here as arising from sedenion automorphisms, and the $S_3$ associated with $\mathrm{Spin}(8)$ triality, which has previously been proposed as a possible origin of three fermion generations.

\begin{table}[h]
\centering
\renewcommand{\arraystretch}{1.5}
\begin{tabular}{|c|c|c|c|c|}
\hline
State & $T_3$ eigenvalue & $Q'$ eigenvalue & $Y$ eigenvalue & Particle \\ \hline

$a_1^{\dagger}a_2^{\dagger}a_3^{\dagger}a_4^{\dagger}f_{+++++}$ 
& $+\frac{1}{2}$ 
& $1$ 
& $1$ 
& $e^+_L$ \\ \hline

$a_i^{\dagger}a_j^{\dagger}f_{+++++}$ 
& $+\frac{1}{2}$ 
& $+\frac{1}{3}$ 
& $-\frac{1}{3}$ 
& $\bar{d}_L^{(\bar{3})}$ \\ \hline

$a_i^{\dagger}a_4^{\dagger}f_{+++++}$ 
& $+\frac{1}{2}$  
& $+\frac{2}{3}$ 
& $+\frac{1}{3}$ 
& $u_L^{(3)}$ \\ \hline

$a_4^{\dagger}a_5^{\dagger}f_{+++++}$ 
& $+\frac{1}{2}$  
& $0$ 
& $-1$ 
& $\nu_L$ \\ 

\hline\hline

$a_1^{\dagger}a_2^{\dagger}a_3^{\dagger}a_4^{\dagger}a_5f_{++++-}$ 
& $-\frac{1}{2}$ 
& $0$ 
& $1$ 
& $\bar{\nu}_L$ \\ \hline

$a_i^{\dagger}a_j^{\dagger}a_5f_{++++-}$ 
& $-\frac{1}{2}$ 
& $-\frac{2}{3}$ 
& $-\frac{1}{3}$ 
& $\bar{u}_L^{(\bar{3})}$ \\ \hline

$a_i^{\dagger}a_4^{\dagger}a_5f_{++++-}$ 
& $-\frac{1}{2}$  
& $-\frac{1}{3}$ 
& $+\frac{1}{3}$ 
& $d_L^{(3)}$ \\ \hline

$a_4^{\dagger}f_{++++-}$ 
& $-\frac{1}{2}$  
& $-1$ 
& $-1$ 
& $e^-_L$ \\ 

\hline\hline

$a_1^{\dagger}a_2^{\dagger}a_3^{\dagger}a_5f_{+++--}$ 
& $0$ 
& $1$ 
& $+2$ 
& $e^+_R$ \\ \hline

$a_i^{\dagger}a_j^{\dagger}a_4a_5f_{+++--}$ 
& $0$ 
& $+\frac{1}{3}$ 
& $+\frac{2}{3}$ 
& $\bar{d}_R^{(\bar{3})}$ \\ \hline

$a_i^{\dagger}a_5f_{+++--}$ 
& $0$  
& $+\frac{2}{3}$ 
& $+\frac{4}{3}$ 
& $u_R^{(3)}$ \\ \hline

$f_{+++--}$ 
& $0$  
& $0$ 
& $0$ 
& $\nu_R$ \\ 

\hline\hline

$a_1^{\dagger}a_2^{\dagger}a_3^{\dagger}f_{+++-+}$ 
& $0$ 
& $0$ 
& $0$ 
& $\bar{\nu}_R$ \\ \hline

$a_i^{\dagger}a_j^{\dagger}a_4f_{+++-+}$ 
& $0$ 
& $-\frac{2}{3}$ 
& $-\frac{4}{3}$ 
& $\bar{u}_R^{(\bar{3})}$ \\ \hline

$a_i^{\dagger}f_{+++-+}$ 
& $0$  
& $-\frac{1}{3}$ 
& $-\frac{2}{3}$ 
& $d_R^{(3)}$ \\ \hline

$a_5^{\dagger}f_{+++-+}$ 
& $0$  
& $-1$ 
& $-2$ 
& $e^-_R$ \\ \hline

\end{tabular}
\caption{Eigenvalues of $T_3$, $Q'$, and $Y$ for the first generation of algebraic states, identified with the first generation of fermions.}
\end{table}

\vspace{-15pt}
\subsubsection*{Acknowledgements}
The author thanks Liam Gourlay for useful discussions and for exploring preliminary aspects of this construction during his PhD work. The present formulation, analysis, and conclusions are solely those of the author. The author is particularly  grateful to Janek Kozicki for his exceptionally careful verification of every calculation in the paper using Mathematica\footnote{The verification notebook is available at \url{https://gitlab.com/cosurgi/verification-of-arxiv-2601.07857}.}.
\appendix
\section{$U(4)$ symmetry generators}\label{sec:appA}

\begin{equation}
\label{SU(4) Generators}
\begin{alignedat}{3} 
    \Lambda_1 &=  -a_2^\dagger{a_1}-a_1^\dagger{a_2}, \hspace{1em} 
    \Lambda_2 &&= ia_2^\dagger{a_1}-ia_1^\dagger{a_2}, \hspace{1em} \Lambda_3 = a_2^\dagger{a_2}-a_1^\dagger{a_1}, \\
    \Lambda_4 &= -a_1^\dagger{a_3}-a_3^\dagger{a_1}, \hspace{1em} 
    \Lambda_5 &&= -ia_1^\dagger{a_3}+ia_3^\dagger{a_1}, \hspace{1em}
    \Lambda_6 = -a_3^\dagger{a_2}-a_2^\dagger{a_3}, \\ 
    \Lambda_7 &= ia_3^\dagger{a_2}-ia_2^\dagger{a_3}, \hspace{1em}
     \Lambda_8 &&= -\frac{1}{\sqrt{3}}(a_1^\dagger{a_1}+a_2^\dagger{a_2}-2a_3^\dagger{a_3}),  \\
         \Lambda_9 &= -a_4^\dagger{a_1}-a_1^\dagger{a_4}, \hspace{1em}    
    \Lambda_{10} &&= ia_4^\dagger{a_1}-ia_1^\dagger{a_4}, \hspace{1em} 
    \Lambda_{11} = -a_4^\dagger{a_2}-a_2^\dagger{a_4}, \\
    \Lambda_{12} &= ia_4^\dagger{a_2}-ia_2^\dagger{a_4}, \hspace{1em} 
    \Lambda_{13} &&= -a_4^\dagger{a_3}-a_3^\dagger{a_4}, \hspace{1em}
    \Lambda_{14} = ia_4^\dagger{a_3}-ia_3^\dagger{a_4}, \\ 
    \Lambda_{15} &= -\frac{1}{\sqrt{6}}(a_1^\dagger{a_1}+a_2^\dagger{a_2}+&&a_3^\dagger{a_3}-3a_4^\dagger{a_4}).
\end{alignedat}
\end{equation}

\begin{eqnarray}
    N_8=\sum_{i=1}^{4} a_i^\dagger a_i
\end{eqnarray}

\bibliographystyle{unsrt}
\bibliography{v3Bibliography}

\end{document}